\documentclass[pre,twocolumn,amssymb,showpacs]{revtex4}

\usepackage{graphicx}
\usepackage{color}
\usepackage{epsfig}
\usepackage{latexsym}
\usepackage{bm}
\usepackage{ulem}

\usepackage{color}

\begin{document}

\renewcommand{\ni}{{\noindent}}
\newcommand{\dprime}{{\prime\prime}}
\newcommand{\be}{\begin{equation}}
\newcommand{\ee}{\end{equation}}
\newcommand{\bea}{\begin{eqnarray}}
\newcommand{\eea}{\end{eqnarray}}
\newcommand{\nn}{\nonumber}
\newcommand{\bk}{{\bf k}}
\newcommand{\bQ}{{\bf Q}}
\newcommand{\q}{{\bf q}}
\newcommand{\s}{{\bf s}}
\newcommand{\bN}{{\bf \nabla}}
\newcommand{\bA}{{\bf A}}
\newcommand{\bE}{{\bf E}}
\newcommand{\bj}{{\bf j}}
\newcommand{\bJ}{{\bf J}}
\newcommand{\bs}{{\bf v}_s}
\newcommand{\bn}{{\bf v}_n}
\newcommand{\bv}{{\bf v}}
\newcommand{\la}{\langle}
\newcommand{\ra}{\rangle}
\newcommand{\dg}{\dagger}
\newcommand{\br}{{\bf{r}}}
\newcommand{\brp}{{\bf{r}^\prime}}
\newcommand{\bq}{{\bf{q}}}
\newcommand{\hx}{\hat{\bf x}}
\newcommand{\hy}{\hat{\bf y}}
\newcommand{\bS}{{\bf S}}
\newcommand{\cU}{{\cal U}}
\newcommand{\cD}{{\cal D}}
\newcommand{\bR}{{\bf R}}
\newcommand{\pll}{\parallel}
\newcommand{\sumr}{\sum_{\vr}}
\newcommand{\cP}{{\cal P}}
\newcommand{\cQ}{{\cal Q}}
\newcommand{\cS}{{\cal S}}
\newcommand{\ua}{\uparrow}
\newcommand{\da}{\downarrow}
\newcommand{\red}{\textcolor {red}}
\newcommand{\blu}{\textcolor {blue}}
\newcommand{\1}{{\oldstylenums{1}}}
\newcommand{\2}{{\oldstylenums{2}}}
\newcommand{\mDelta}{\varepsilon}
\newcommand{\m}{\tilde m}
\def\lsim {\protect \raisebox{-0.75ex}[-1.5ex]{$\;\stackrel{<}{\sim}\;$}}
\def\gsim {\protect \raisebox{-0.75ex}[-1.5ex]{$\;\stackrel{>}{\sim}\;$}}
\def\lsimeq {\protect \raisebox{-0.75ex}[-1.5ex]{$\;\stackrel{<}{\simeq}\;$}}
\def\gsimeq {\protect \raisebox{-0.75ex}[-1.5ex]{$\;\stackrel{>}{\simeq}\;$}}

\title{Additivity property and emergence of power laws in 
nonequilibrium steady states}

\author{Arghya Das$^{1}$, Sayani Chatterjee$^{1}$} 
\author{Punyabrata Pradhan$^{1}$} 
\email{punyabrata.pradhan@bose.res.in}
\author{P. K. Mohanty$^{2,3}$}

\affiliation{ $^1$Department of Theoretical Sciences, S. N. Bose National Centre for Basic Sciences, Block-JD, Sector-III, Salt Lake, Kolkata 700098, India \\
$^2$CMP Division, Saha Institute of Nuclear Physics, 1/AF Bidhan
Nagar, Kolkata 700064, India \\ $^3$Max Planck Institute for the
Physics of Complex Systems, 01187 Dresden, Germany}

\begin{abstract}

\noindent{   
We show that an equilibrium-like additivity property can remarkably lead to power-law distributions observed frequently in a wide class of out-of-equilibrium systems. The additivity property can determine the full scaling form of the distribution functions and the associated exponents. The asymptotic behaviour of these
distributions is solely governed by branch-cut singularity in the variance of subsystem mass. To 
substantiate these claims, we explicitly calculate, using the 
additivity property, subsystem mass distributions in 
a wide class of previously studied mass aggregation models as well 
as in their variants. These results could help in thermodynamic 
characterization of nonequilibrium critical phenomena.
}

\typeout{polish abstract}

\end{abstract}


\maketitle

\section{Introduction}

Simple power-law scaling is ubiquitous in nature 
\cite{BakBook}. They appear in the distributions of drainage area 
of rivers \cite{Scheidegger}, droplet size \cite{droplets1, 
droplets2}, size of clusters formed in polymerization processes 
\cite{Ziff_PRL1982}, rain size \cite{rain}, size of fragments in 
fractured solids \cite{Hermann}, population and wealth 
\cite{population, wealth}, and in stock market fluctuations 
\cite{Stanley}, etc. Evidently, power laws, which are usually 
associated with criticality  through emergence of a diverging 
length scale, are observed in widely unrelated systems, suggesting 
existence of some broad underlying principle. Recent evidence that 
living systems might be operating, independent of most of the 
microscopic details, in the vicinity of a critical regime 
\cite{Bio} indeed invoke further questions - how and why systems 
adapt to near-criticality.

There have been several attempts to reveal the origin of the power 
laws in nature, through studies of paradigmatic nonequilibrium 
models - most appealing being sandpile \cite{BTW1987, 
DickmanPRL1998, MohantyPRL2002} and mass aggregation models 
\cite{Vigil, Krapivsky, TakayasuPRL1989, Majumdar_PRL1993, 
Barma_PRL1998, Puri}. Many of these models - where there 
is a conservation law or, in case of violation, the law is weakly 
violated in the sense that the systems are slowly driven - are 
intimately connected to each other. For example, the mass 
aggregation models \cite{TakayasuPRL1989, Majumdar_PRL1993, 
Barma_PRL1998, Barma_JSP2000} are connected to directed abelian 
sandpile model \cite{Dhar} or to the models of river network 
\cite{Scheidegger}.

In this paper, we argue that power-law distributions in 
out-of-equilibrium systems can arise simply from additivity 
property, the tenet of equilibrium thermodynamics. We find that the 
divergence in the response function is the key: Diverging 
fluctuations can, in principle, arise from distributions other than 
power laws, which are however prohibited if one imposes additivity 
and consequent fluctuation-response (FR) relation. The response 
function determines the full scaling form of the distribution, {\it 
at as well as away} from criticality, and critical exponents 
originate from the singularity in the response function. To 
demonstrate this, we consider  mass aggregation models which are 
known to have nonequilibrium steady state with scale invariant 
structures. At {\it all} mass densities, the distribution function 
$P_v(m)$ of mass $m$ in a subsystem of volume $v$, which is 
obtained solely from the FR relation, is shown to have a scaling 
form $P_v(m) \sim m^{-\tau} \exp (\tilde \mu m)$. The quantity 
$\tilde \mu(\rho) = \mu(\rho) - \mu(\rho_c)$, inverse of a 
cut-off mass $m^*(\rho) = - 1/\tilde \mu(\rho)$, is an analogue of 
equilibrium-like chemical potential and provides a useful 
thermodynamic interpretation of the emergence of power laws in 
nonequilibrium steady states. The exponent $\tau$ and 
the critical properties of chemical $\mu(\rho)$ arise from a multiple-pole 
or branch-cut singularity in the variance at a critical mass 
density $\rho_c$. As the critical density is approached $\rho 
\rightarrow \rho_c$, nonequilibrium chemical potential vanishes 
$\mu(\rho) \rightarrow 0$, leading to pure power laws. Beyond the 
critical density $\rho > \rho_c$, there is a gas-liquid like phase 
coexistence.

The above result immediately provides answer to why $m^{-5/2}$ 
power law, at or away from criticality, appears so often in mass 
aggregation models - especially in higher dimensions, at all 
densities and irrespective of that the motion of the diffusing 
masses is biased or not \cite{Barma_PRL1998, Barma_JSP2000, 
Rajesh_PRE2001, Rajesh_PRE2002}. Interestingly, the same power law 
appears in $k$-mer distribution in the classic Flory-Stockmayer 
\cite{Stockmayer} theory of polymerization and also in particle 
number distribution in three dimensional ideal Bose gas near 
critical point, irrespective of whether the systems are in or out 
of equilibrium - thus indicating a {\it universality}. We 
demonstrate that the $m^{-5/2}$ law is a consequence of a 
simple-pole singularity in the variance. The whole analysis is 
extended 
also to nonconserved mass aggregation models. We validate our 
theory by explicitly calculating mass distributions in previously 
studied mass aggregation models and their variants and by comparing 
them with simulations.

Organization of the paper is as follows. In section II.A, we discuss additivity property; in section II.B, we discuss the connection between singularity in the variance and the asymptotic behaviour of the mass distribution function. In section III, we illustrate our analytic methods in a broad class of model systems, both in conserved-mass aggregation models and its nonconserved versions. Finally, we summarize our results with a concluding perspective.

\section{Theory}

\subsection{Additivity Property}

We start by invoking an additivity property which
a wide class of systems, irrespective of that they are in or out of 
equilibrium, could possess. Consider a continuous-mass 
(generalization to discrete masses is straightforward) transport 
process on a lattice of size $V$ and then divide the system in 
$\nu=V/v$ identical subsystems or cells, $k$th subsystem with mass $m_k$, where total mass $M=\sum_k m_k$ remains conserved. Provided the subsystems are large compared to spatial correlation length, they could be considered statistically almost independent \cite{Eyink1996, Bertin_PRL2006, Bertin_PRE2007, PRL2010}. 
In that case, the joint subsystem mass distribution in 
the steady state can be written in a product form, \be {\cal
P}[\{m_k\}] \simeq \frac{\prod_{k=1}^{\nu} w_v(m_k)}{Z(M,V)}
\delta\left( \sum_k m_k - M \right), \label{additivity1} \ee where
$w_v(m_k)$ is an unknown weight factor (to be determined later; see 
Eq. \ref{Pm2}) depending only on the subsystem mass 
$m_k$, $Z=\prod_k \int dm_k w_v(m_k) \delta(\sum_k m_k -M) \equiv 
\exp[-Vf(\rho)]$ the partition sum, $f(\rho)$ a nonequilibrium free
energy density and $\rho=M/V$ mass density. The product form in 
Eq. \ref{additivity1} amounts to an equilibrium-like additivity 
property, in the sense that a free energy function $F 
= \sum_k \ln w_v(m_k)$ is minimized in the macrostate.

Using standard statistical mechanics \cite{Kardar}, Eq. 
\ref{additivity1} leads to the probability distribution 
${\rm Prob} [m_k \in (m,m+dm)]=P_v(m)dm$ for subsystem mass where 
\be
P_v(m) = \frac{w_v(m) e^{\mu m}}{\cal Z}
\label{Pm1}
\ee 
with $\mu(\rho)$ a nonequilibrium chemical potential, and ${\cal 
Z}$ the normalization constant. The weight factor $w_v(m)$ and 
chemical potential $\mu(\rho)=df/d\rho$ can be obtained using a 
fluctuation-response relation \cite{Eyink1996, Bertin_PRL2006, 
Bertin_PRE2007, PRL2010, 
PRL2014}, \be \frac{d\rho}{d\mu} = \sigma^2(\rho), \label{FR1} 
\ee where the scaled variance $\sigma^2(\rho) = (1/v) (\langle 
m_k^2\rangle - \langle m_k \rangle^2)$ in the limit of $v \gg 1$. The free energy density 
function $f(\rho)$ can be obtained through the relation 
$\mu(\rho) = {d f}/ {d\rho}$, i.e., $f(\rho) = \int \mu(\rho) d\rho 
+ \beta $ with chemical potential $\mu(\rho) = \int 1/ 
\sigma^2(\rho) d\rho + \alpha$ (obtained from Eq. \ref{FR1}) and 
$\alpha$ and $\beta$ arbitrary constants of integration. Then 
Laplace transform of $w_v(m)$ is written as 
$\tilde{w}_v(s) = \int w_v(m) \exp(-s m) dm \equiv 
e^{-\lambda_v(s)}$, i.e.,
\be 
e^{-\lambda_v(s)} = \int w_v(m) e^{-s m} dm. \label{L-transform}
\ee 
Then, the function $\lambda_v(s)$ can be obtained from Legendre transform of 
free energy density function $f(\rho)$ \cite{Touchette}, 
\be 
\lambda_v(s) = v \left[ {\bf inf}_{\rho} 
\{ f(\rho) + s \rho \} \right] = v [f(\rho^*) + s \rho^*], 
\label{s1} 
\ee 
where $\rho^*(s)$ is the solution of 
\be
s = - \mu (\rho^*).
\label{rho*}
\ee 
As discussed later, Eq. \ref{s1} requires concavity and differentiability of $f(\rho)$. In the discrete case, the weight factor can be calculated as $w_v(m)=({1}/{2\pi i}) \int_C {\tilde w_v (z)}/{z^{m+1}} dz$ where $\tilde w_v(z) = \sum_{m=0}^{\infty} z^m w_v(m)$ is obtained 
from $\tilde w(s)$ by substituting $s = -\ln z$ with $C$ a suitably
chosen contour in the complex $z$-plane.

\subsection{Singularity in Variance and Mass Distribution}

Importantly, the distribution function $P_v(m)$ is determined 
solely by the functional form of the scaled variance $\sigma^2 
(\rho)$. We argue below that singular response functions generate 
only power-law distributions. Other functional form of mass distribution $P_v(m)$ 
with diverging variance is also possible \cite{note}, which, we show, however are not allowed if the FR relation holds. In this paper, we mainly focus on multi-pole singularity at a finite density $\rho_c$,
\bea \sigma^2(\rho) =  \left\{
\begin{array} {cc}
    \frac{g(\rho)}{(\rho_c-\rho)^n} &\rm{~~~for~\rho < \rho_c,}  \cr
   \infty & \rm{~~~otherwise.}
\end{array} \right. \label{sigma1}
\eea 
This form, with $0<n<\infty$, is relevant in the context of a wide class of mass aggregation models as discussed in section III. The analytic part $g(\rho)$ is not 
particularly relevant in determining the asymptotic form  of the 
distribution $P_v(m)$, however it contributes to the exact form of 
$P_v(m)$ (discussed in section II.B.IV). In fact, other kinds of singularities, such as logarithmic singularity $\sigma^2 (\rho) \sim  [\ln(\rho_c - \rho)]^p$ or exponential singularity $\exp[(\rho_c - \rho)^{-p}]$ where $p>0$, $1/|\rho - \rho_c|^n$, and the case with $n<0$ can also arise. One can show that they all lead to power laws, possibly with logarithmic corrections to the power-law scaling (discussed in the following sections).

\begin{figure}
\begin{center}
\leavevmode
\includegraphics[width=8.5cm,angle=0]{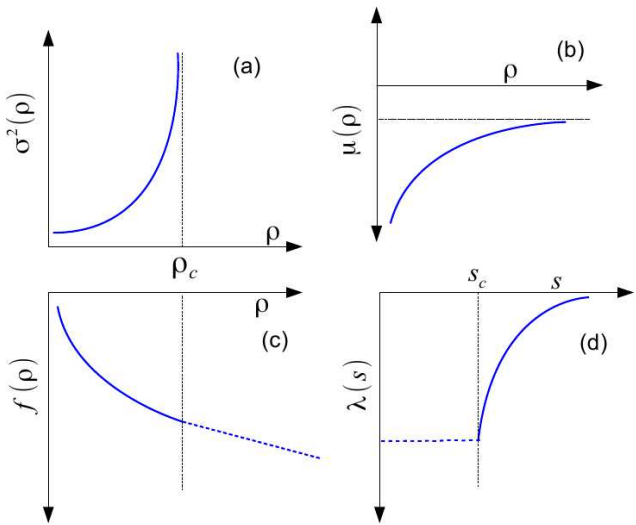}
\caption{(Color online) Schematic representation of condensation transition. 
Panel (a): variance $\sigma^2(\rho)$ as a function of density $\rho$; panel (b): chemical potential $\mu(\rho)$ as a function of $\rho$; panel (c): free energy density $f(\rho)$ as a function of $\rho$; panel (d): Legendre transform $\lambda(s)$ of free energy density as a function of $s$.} \label{mu-f-l}
\end{center}
\end{figure}

The divergence in the variance, as in Eq. \ref{sigma1} (or in the cases of logarithmic and exponential divergence), indeed has broad 
implications, not only in conserved-mass aggregation models but 
also in their nonconserved versions. Note that the FR relation in 
Eq. \ref{FR1} implies that free
energy density $f(\rho)$ is not a strictly concave function of
$\rho$ and has a linear branch of slope $\mu(\rho_c)$ for
$\rho \ge \rho_c$. Moreover, $f''(\rho=\rho_c)=\mu'(\rho=\rho_c)=0$
(prime denotes derivative w.r.t. $\rho$) implies a point of
inflection in $f-\rho$ curve at $\rho = \rho_c$. That is, free energy density function can be written as
\bea f(x) =  \left\{
\begin{array} {cc}
    \int \mu(x) dx \mbox{~~~~~~~~~~~~~~~~~~~for~} x < \rho_c, \cr
   \mu(\rho_c)(x-\rho_c) + f(\rho_c)  \mbox{~~~otherwise.}
\end{array} \right. \label{f-construction}
\eea 
Consequently, Legendre transform of $f(\rho)$ develops a 
branch-cut singularity (see Eq. \ref{lambda-s}); for schematic 
representation of the above analysis, 
see Fig. \ref{mu-f-l}. This construction of 
a nonequilibrium free energy function $f(\rho)$ from a general 
thermodynamic consideration readily explains the phase coexistence 
between a fluid and a condensate, as observed in the past in many 
out-of-equilibrium systems (discussed in section III).

\subsubsection{Multi-pole singularity}

To analyse the behaviour of $\lambda_v(s)$ in the case of Eq. \ref{sigma1}, we integrate Eq.
\ref{FR1} near $\rho=\rho_c$ and obtain \be \mu (\rho) \simeq -
\frac{(\rho_c-\rho)^{n+1}}{(n+1)g(\rho_c)} \left[ 1+ {\cal O}(\rho
-\rho_c) \right] + \alpha, \label{mu-s} \ee which gives 
$(\rho_c - \rho^*) \simeq [(n+1)g(\rho_c)(s+\alpha)]^{1/(1+n)}$ by using 
Eqs. \ref{rho*} and \ref{mu-s}. Integrating chemical potential $\mu(\rho)$, we get free energy function \be 
f(\rho) \simeq \frac {(\rho_c - \rho)^{n+2}}{(n+1)(n+2)g(\rho_c)} + 
\alpha \rho + \beta \ee and write $\lambda_v(s) = v[f(\rho^*) + s 
\rho^*$], in leading order, as \be \lambda_v(s) \simeq v[a_0 + a_1 
(s+\alpha) + a_2(s + \alpha)^{\frac{n+2}{n+1}}], \label{lambda-s} 
\ee where $a_0, a_1, a_2$ are constants. Thus, we obtain $\tilde 
w(s) = \exp[-\lambda_v(s)] \simeq {\rm const.}\times [1 - v a_1 (s 
+ \alpha) - v a_2 (s+\alpha)^{1+1/(1+n)}]$ in leading orders of $(s+\alpha)$, implying $$w_v(m) \sim 
\frac{e^{-\alpha m}}{m^{\tau}},$$ 
where, for large subsystem masses $m \gg v$, the power-law exponent $\tau$ in the denominator is given by 
\be 
\tau = \left[ 2+\frac{1}{(1+n)} \right], 
\label{tau}
\ee with the following inequality $2< \tau <3$ (since $0 < n < \infty$ in Eq. \ref{sigma1}). This translates  into the mass distribution having a scaling form,
\be 
P_v(m) \propto \frac{1}
{m^{\tau}} e^{\tilde \mu (\rho) m} \equiv \frac{1}{(m^*)^\tau} \Phi\left( \frac{m}{m^*} \right), 
\label{Pm2} 
\ee 
where $\tilde 
\mu(\rho) = \int_{\rho_c}^{\rho} 1/\sigma^2(\rho) d\rho = \mu(\rho) 
- \mu(\rho_c)$ is an effective chemical potential, inverse of which gives a 
cut-off $m^* = -1/\tilde \mu$ in the distribution, and the scaling function $\Phi(x)=x^{-\tau} \exp(- x)$. Later, we 
explicitly calculate $\tilde \mu (\rho)$ in specific model systems. Note that $\tilde \mu (\rho_c)=0$ at {\it critical point $\rho = 
\rho_c$} and consequently $P_v(m)$ becomes a pure power law. 
Moreover, by defining a critical exponent $\delta=1+n$ as $\tilde 
\mu(\rho) \sim (\rho_c - \rho)^{\delta}$, we get a scaling 
relation $\delta(\tau-2) = 1$.

\subsubsection{Logarithmic singularity}

Now, we consider the case of logarithmic singularity where variance $\sigma^2(\rho)$ diverges logarithmically as given below,
\bea \sigma^2(\rho) =  \left\{
\begin{array} {cc}
    {g(\rho) [\ln(\rho_c - \rho)]^p} &\rm{~~~for~\rho < \rho_c,}  \cr
   \infty & \rm{~~~otherwise.}
\end{array} \right. \label{sigma2}
\eea 
Integrating Eq. \ref{FR1} near $\rho = \rho_c$, we obtain chemical potential, in leading order of $(\rho_c - \rho)$,
\be 
\mu(\rho) \simeq - \frac{(\rho_c - \rho)}{g(\rho_c) [\ln(\rho_c - \rho)]^p} + \alpha,
\ee
which gives $(\rho_c - \rho^*) \simeq g(\rho_c) [\ln(\rho_c - \rho^*)]^p (s+\alpha)$ from Eq. \ref{rho*}. Free energy density is obtained by integrating above chemical potential $\mu(\rho)$,
\be 
f(\rho) \simeq - \frac{(\rho_c - \rho)^2}{2g(\rho_c) [\ln(\rho_c - \rho)]^p} + \alpha \rho + \beta,
\ee 
and, accordingly, its Legendre transform, in leading orders,
\be \lambda_v(s) \simeq v[a_0 + a_1 
(s+\alpha) + a_2(s + \alpha)^2 \{\ln (s+\alpha)\}^p]. \label{lambda-s-log} 
\ee
For large mass $m \gg v$, this implies that the weight factor has a functional form of a power law with logarithmic correction, $$w_v(m) \sim \frac{(\ln m)^{p-1}}{m^3} e^{-\alpha m},$$ and the corresponding mass distribution function,
\be 
P_v(m) \propto \frac{(\ln m)^{p-1}}{m^3} e^{\tilde \mu (\rho) m}, \label{Pm-log} 
\ee
where effective chemical potential $\tilde \mu(\rho) = \mu(\rho) - \mu(\rho_c)$ and the power-law exponent in the denominator is $\tau=3$, the borderline case of Eqs. \ref{sigma1} and \ref{tau} with $n=0$.

\subsubsection{Exponential singularity}

We also consider the case where the variance diverges exponentially, $\sigma^2(\rho) \sim \exp(\rho_c - \rho)^{-p}$ for $\rho < \rho_c$ and $\sigma^2(\rho) = \infty$ otherwise. The analysis is similar to the ones given above. Substituting chemical potential $\mu(\rho) \simeq {\rm const} \times (\rho_c - \rho)^{p+1} \exp [-(\rho_c-\rho)^{-p}] + \alpha$ in Eq. \ref{rho*} and solving in leading order of $(s+\alpha)$, we get $(\rho_c-\rho^*) \sim \{\ln(s+\alpha)\}^{-1/p}$ and consequently
\be 
\lambda_v(s) \simeq v \left[ a_0 + a_1 (s+\alpha) + a_2 (s+\alpha) \{\ln(s+\alpha)\}^{-1/p} \right]. \label{lambda-s-exp} 
\ee
For large mass $m \gg v$, this leads to the mass distribution function
\be 
P_v(m) \propto \frac{(\ln m)^{-1-1/p}}{m^2} e^{\tilde \mu (\rho) m}, \label{Pm-exp} 
\ee
where $\tilde \mu(\rho) = \mu(\rho) - \mu(\rho_c)$ and the 
power-law exponent in the denominator is $\tau=2$, the borderline case of Eqs. \ref{sigma1} and \ref{tau} with $n= \infty$.

\subsubsection{Subsystem mass distribution}

For any finite $v$, it is not easy to find the distribution function $P_v(m)$ at small or intermediate values of masses $m \sim v$ because, in that case, one requires to invert Eq. \ref{L-transform} using inverse Laplace transform, i.e., by evaluating the integral,
\be 
w_v(m) = \frac{1}{2\pi i} \int_C e^{-\lambda_v(s) + ms} ds,
\label{inv-L-transform}
\ee 
on the complex $s$-plane; the contour $C$ on the complex plane should be chosen such that the integral converges. However, in the models we consider here, it is not possible to get an exact functional form of $\lambda_v(s)$ for all $s$, which actually involves solving the transcendental Eq. \ref{rho*}.  However, for large subsystem size $v \gg 1$, the analysis simplies as the function $-(1/v) \ln w_v(m)$ is simply related to $\lambda_v(s)/v$ by Legendre transformation \cite{Touchette} and therefore, in leading order of $m$, is the free energy density function $f(m/v)$ itself, which has been already constructed in Eq. \ref{f-construction} (see Fig. \ref{mu-f-l} and related discussions). The subsystem mass distribution function now can be written as
\be
P_v(m) \propto \frac{e^{-vf(m/v)+\mu(\rho) m}}{(a+m)^{2.5}},
\label{Pvm}
\ee
where the denominator is essentially a sub-leading correction to the free energy function, with $a \sim {\cal O}(v)$ being a model-dependent cut-off mass. The correction term is obtained from the fact that, for large masses $m \gg v$, free energy function $f(x)$ has a linear branch (see Eq. \ref{f-construction}) and the mass distribution function must have the asymptotic form $P_v(m) \sim m^{-\tau} \exp(\tilde \mu m)$ as in Eq. \ref{Pm2}.

\section{Models and illustration}

\subsection{Conserved Mass Aggregation Models (CMAM)}

We now substantiate the above claims in a broad class of nonequilibrium models which were studied intensively in the last couple of decades. In this paper, we mainly focus on the models having multi-pole singularity in the variance. Other singularities, e.g., exponential or logarithmic, are possible, however not as common as the multi-pole one. For 
the purpose of illustrations, we specifically consider the case 
with $n=1$ and mass distribution at a single site, i.e., $v=1$. 
We define a generalized version of conserved mass aggregation 
models (CMAM) studied in \cite{Barma_PRL1998, Barma_JSP2000, Puri, 
Barma_PRE2000}, for simplicity on a one dimensional lattice of $L$
sites. We mainly focus on symmetric mass transfer, i.e., masses are transferred symmetrically to either of the neighbours.
Here masses (or particles) diffuse, fragment and coalesce
stochastically with either of the nearest-neighbour masses according to the
following dynamical rules: (1) diffusion of mass $m_i$ at site $i$
to $i\pm1$ with mass-dependent rate $D(m_i)$ where $m_i
\rightarrow 0$ and $m_{i\pm1} \rightarrow m_{i\pm1}+m_i \nonumber$
and (2) fragmentation of a discrete mass $\Delta$ at site $i$,
provided $\Delta \le m_i$, and coalescence of the mass to either
of the sites $i\pm1$ with mass-dependent rate $w(\Delta)$ where
$m_i \rightarrow m_i-\Delta$ and $m_{i\pm1} \rightarrow
m_{i\pm1}+\Delta$ with $\Delta=1,2, \dots$ (continuous $\Delta$ 
considered later). Total mass
$M=\sum_{i=1}^{V} m_i$ is conserved in this process. Even for
these simple dynamical rules, the steady state weight in general
is not exactly known. However, as spatial correlations are small,
the additivity property as in Eq. \ref{additivity1}, to a good
approximation, is expected to hold.

\begin{figure}
\begin{center}
\leavevmode
\includegraphics[width=8.0cm,angle=0]{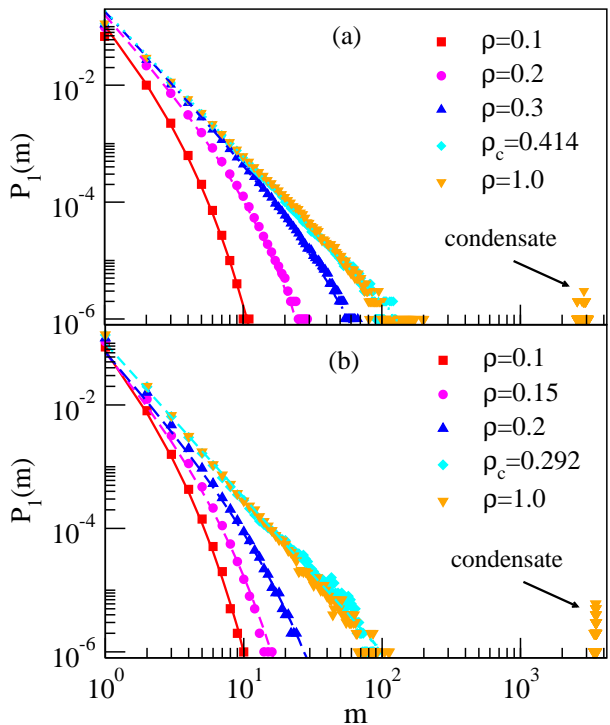}
\caption{(Color online) Single-site ($v=1$) mass distribution functions $P_1(m)$ (points - simulations) in conserved-mass aggregation models (CMAMs) is compared with analytic expression in Eq. \ref{Pm2} (lines - theory) for various densities. Panel (a): mass chipping rates $w_1=1$, $w_2=0$; panel (b): mass chipping rates $w_1=0$, $w_2=1$. In all cases, mass diffusion rate $D=1$ and system size $L=5000$.}
\label{CMAM}
\end{center}
\end{figure}

We calculate the variance $\sigma^2 (\rho)$ of mass at a single 
site in various special cases, using the additivity property 
Eq. \ref{additivity1} with $v=1$. We take diffusion rate
$D(m_i)=1$, independent of mass $m_i$, $w(\Delta=1) = w_1$
(rate of single particle chipping), $w(\Delta=m_i-1)=w_2$ (rate of
all-but-one particle chipping) and $w(\Delta)=0$ otherwise.

\subsubsection{Case I: CMAM with $w_1=1, w_2=0$}

For $w_1=1$ and $w_2=0$ and symmetric mass transfer, the model 
becomes the symmetric one studied in \cite{Barma_PRL1998, 
Barma_JSP2000} (our model is a variant of those studied in \cite{Vigil, Krapivsky}). For $\rho \le \rho_c$, using additivity 
property, we exactly calculated the variance and consequently 
chemical potential with the critical density $\rho_c=\sqrt{2}-1$
We can calculate the variance as given below (for details, see Appendix A)
\bea
\sigma^2(\rho) &=& \frac{\rho(1+\rho)(1+\rho^2)}{(1-2\rho-\rho^2)} \nonumber \\
&=& \frac{\rho(1+\rho)(1+\rho^2)}{(\rho_c-\rho)(\sqrt{2}+1+\rho)},
\label{Barma}
\eea 
with $\rho_c=(\sqrt{2}-1),$
for which one can obtain a chemical potential $\mu(\rho)$ and free energy function $f(\rho)$, by integrating the fluctuation-response relation as in Eq.2 in the main text,
\bea
\mu(\rho) = \int \frac{1}{\sigma^2(\rho)} d\rho 
= - 2 \tan^{-1} \rho + \ln \left( \frac{\rho}{1+\rho} \right) + \alpha 
\eea
and, upon one more integration,
\bea
f(\rho) = \int \mu(\rho) d\rho  = - 2 \rho \tan^{-1} \rho + \rho \ln \left( \frac{\rho}{1+\rho} \right)
\nonumber \\ 
- \ln \left( \frac{1+\rho}{1+\rho^2} \right) 
+ \alpha \rho + \beta
\eea
where $\alpha$ and $\beta$ are two arbitrary constants of integration. For large mass $m \gg 1$, the mass 
distribution function is calculated to be $P_1(m) \propto 
m^{-5/2}\exp[(\mu(\rho)-\mu(\rho_c))m]$. (for details,
see Appendix B).

In panel (a) of Fig. \ref{CMAM}, we have plotted single-site ($v=1$) mass distribution function $P_1(m)$, obtained from simulations, as a function of mass $m$ for various values of
densities $\rho=0.1$, $0.2$, $0.3$, $0.414$ and $1.0$ and compare them with the theoretical expression in Eq. \ref{Pm2}. The theoretical results have a quite good agreement with the simulation results, especially at large mass $m\gg 1$. In panel (a) of Fig. \ref{CMAM-Pvm}, we have plotted subsystem mass distribution function $P_v(m)$, with $v=100$, for densities $\rho=0.1$ and $0.2$ and compare them  with the theoretical expression in Eq. \ref{Pvm} with cut-off mass $a \approx 20$; agreement between simulations and theory is reasonably good.

\subsubsection{Case II: CMAM with $w_1=0, w_2=1$}

The CMAM with $w_1=0, w_2=1$ is a variant of the models studied in \cite{Puri}. In this case, for $\rho \le \rho_c$, the 
variance and chemical potential can be exactly obtained using 
additivity property (for details, see Appendix A). The variance is given by
\bea
\sigma^2(\rho) &=& \frac{\rho(1-\rho)(2\rho^2-2\rho+1)}{2\rho^2-4\rho+1}
\nonumber \\
&=&  \frac{\rho(1-\rho)(2\rho^2-2\rho+1)}{(\rho_c-\rho)(2+\sqrt{2}-2\rho)}.
\eea
There is a simple pole at the critical density $\rho_c = 1 - {1}/{\sqrt{2}}$. By integrating fluctuation-response relation Eq. \ref{FR1}, we get chemical potential
\begin{equation}
\mu(\rho) = 2 \tan^{-1}(1-2\rho)-\ln \left[ \frac{1}{2\rho(1-\rho)}-1\right] + \alpha,
\end{equation}
and then free energy density
\bea
f(\rho)=2\rho \tan^{-1}(1-2\rho)-\ln(1-\rho)+\ln(1-2\rho+2\rho^{2})
\nonumber \\
-\rho \ln \left[ \frac{1}{2\rho(1-\rho)}- 1 \right] + \alpha \rho + \beta ~~.
\eea
In panel (b) of Fig. \ref{CMAM}, we have plotted single-site ($v=1$) mass distribution function $P_1(m)$, obtained from simulations, for various values of densities $\rho=0.1$, $0.15$, $0.2$, $0.292$ and $1.0$. We find that the simulation results agree remarkably well with the analytical scaling form
$P_1(m) \propto m^{-5/2} \exp[(\mu(\rho)- \mu(\rho_c))m]$ as in Eq. \ref{Pm2} with $\tau=5/2$ (for details of the derivation, see Appendix B). In panel (b) of Fig. \ref{CMAM-Pvm}, we have plotted subsystem mass distribution function $P_v(m)$ for $v=100$ for densities $\rho=0.1$ and $0.15$ and compared them with theory, Eq. \ref{Pvm} with $a \approx 25$; agreement between simulations and theory is reasonably good.

\begin{figure}
\begin{center}
\leavevmode
\includegraphics[width=8.0cm,angle=0]{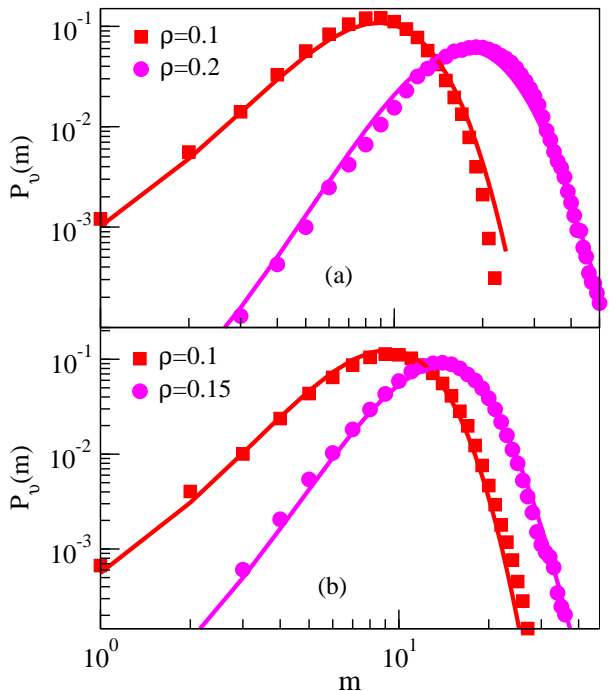}
\caption{(Color online) Subsystem mass distribution functions $P_v(m)$ (points - simulations) in conserved mass aggregation models (CMAMs) is compared with analytic expression in Eq. \ref{Pvm} (lines - theory) for various densities. Panel (a): mass chipping rates $w_1=1$, $w_2=0$ and $a \approx 20$; panel (b): mass chipping rates $w_1=0$, $w_2=1$ and cut-off mass $a \approx 25$. In all cases, mass diffusion rate $D=1$, system size $L=10^5$ and subsystem size $v=100$.}
\label{CMAM-Pvm}
\end{center}
\end{figure}

\subsubsection{Other Cases}

\begin{figure}
\begin{center}
\leavevmode
\includegraphics[width=8.5cm,angle=0]{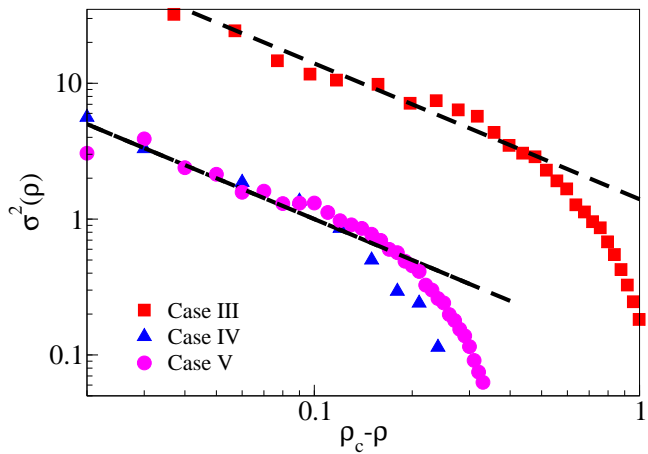}
\caption{(Color online) Variance $\sigma^2(\rho)$ {\it vs.} $(\rho_c-\rho)$. Black line
is ${\rm const.} \times (\rho_c-\rho)^{-n}$ with $n=1$. Red 
rectangles are for Case III (one and two particle fragmentation), 
blue triangles are for Case IV [$\Delta=1,2, \dots$ is discrete 
with fragmentation rate $w(\Delta)= \exp(-\Delta)$] and red circles 
are for case V [$\Delta>0$ is continuous with fragmentation rate 
$w(\Delta) = \exp(-\Delta)$]. Diffusion rate $D(m)=1$ throughout.}
\label{rhoc}
\end{center}
\end{figure}

We have also studied, through simulations, various other cases 
(with $D=1$): Case III. - $w(\Delta=1)=w_1$, $w(\Delta=2)=w_2$ 
and $w(\Delta)=0$ otherwise, Case IV. - a discrete-mass model with 
$w(\Delta)=\exp(-\Delta)$ and Case V. - a continuous-mass model 
with $w(\Delta)=\exp (-\Delta)$. In these cases, in the absence of 
an analytical expression of $\sigma^2(\rho)$, we checked in 
simulations (see Fig. \ref{rhoc}) that the variance near critical point indeed has the 
behaviour $\sigma^2 \sim (\rho_c-\rho)^{-n}$, with $n=1$, which therefore leads to the same power-law 
exponent $\tau=5/2$.

One can also define an asymmetric version of the CMAMs discussed above. In one dimension, there are some nontrivial spatial correlations and the above mean-field analysis fails to capture the mass fluctuations in the system. However, in higher dimensions, the above results qualitatively remain same also for the asymmetric mass transfer and is consistent with \cite{Rajesh_PRE2002}.

Interestingly, the exponent $\tau=5/2$ 
appears also in the distribution of particle numbers in ideal 
Bose gas in three dimensions ($3$D) near the critical point where 
Bose-Einstein condensation (BEC) occurs. This could be easily 
understood 
from the fact that particle-number fluctuation in the case of $3$D 
Bose gas has the same critical behaviour $\sigma^2(\rho) \sim 
(\rho_c-\rho)^{-n}$, with $n=1$, as in these `mean-field' 
nonequilibrium systems having negligible spatial correlations. 
That, on a mean-field level, the nonequilibrium aggregation models 
belong to the universality class of equilibrium Bose gas in $3$D, 
so far has not been realized.

It is quite instructive to consider a limiting case of Eq. \ref{sigma1} with $n=0$, $\rho_c = \infty$ and $g(\rho) \sim \rho^{1-\delta}$ at large density, i.e., the variance $\sigma^2 (\rho) \sim \rho^{1-\delta}$, with $\delta < -1$, diverges algebraically at infinite density. As there is no singularity in the variance at any finite density, our analysis quite straightforwardly shows that condensation transition cannot occur, consistent with the past observations in the CMAM with mass-dependent diffusion \cite{Rajesh_Das_PRE2002}. Asymptotic scaling of the mass distribution can be obtained as follows. Using Eq. \ref{FR1}, we get $\mu(\rho) \sim \rho^{\delta}$ (setting $\alpha = 0$ without loss of any generality) and, consequently, Laplace transform of weight factor $w_v(m)$,
\be
\tilde w_v(s) \simeq a_0 + a_1 s^{1+1/\delta},
\ee 
immediately leading to mass 
distributions having a scaling form $P_v(m) \propto m^{-\tau} \exp(\tilde \mu m) \equiv (m^*)^{-\tau} \Phi(m/m^*)$. Here, the scaling function $\Phi(x) = x^{-\tau} \exp(-x)$ with $m^* \sim \rho^{-\delta}$ and power law exponent $\tau = 2 + 1/\delta$  with $1< \tau < 2$ (as $\delta<-1$), leading to a relation $\delta (\tau-2) = 1$. The scaling form was numerically observed in 
\cite{Rajesh_Das_PRE2002}. Interestingly, the borderline case with $\delta = -1$ generates gamma distributions, which are found in a broad class of mass transport processes \cite{PRL2014} and have been also observed in a limiting case of conserved-mass aggregation models studied in Ref. \cite{Rajesh_Das_PRE2002}.

\subsection{Nonconserved Mass Aggregation Models}

In this section, we discuss a nonconserved version of the mass aggregation models where systems can exchange mass, though {\it weakly}, with environment. In this case, in addition to 
the earlier defined two processes (1) diffusion and (2) fragmentation of masses, a 
particle now can be adsorbed with rate $q$ and desorbed at a site 
with rate $p$, provided the site is occupied, where $p, q 
\rightarrow 0$ (i.e., weak exchange) with the ratio $r = q/p$ finite. Due to adsorption 
and desorption processes, total mass in the system is not 
conserved. This model is related to several models studied in the 
past for finite $p$ and $q$ \cite{Barma_PRL1998, TakayasuPRL1989, 
Barma_PRE2000, Rajesh_PRE2004}. Interestingly, in the limit of $p, 
q \rightarrow 0$, mass fluctuation in a nonconserved model can be 
obtained from the occupation probability of a site in its conserved 
version (i.e., $p = q = 0$) \cite{Mukamel_PRL2012, PRE2015}. 
Let us first define, in the space of total 
mass $M$, a transition rate $T_{M+1,M}$ from mass $M$ to $M+1$. In 
the steady state, the probability $P(M)$ that the system has mass 
$M$ satisfies a balance condition $$P(M) T_{M+1,M} = P(M+1) T_{M, 
M+1}$$ where the mass distribution $P(M)$ can be obtained as \be 
P(M+1) = \left[ \prod_{M=0}^M \frac{T(M \rightarrow M+1)}{T(M+1 
\rightarrow M)} \right] P(0). \ee As the ratio of transition rates 
can be written as $$\frac{T_{M+1,M}}{T_{M, M+1}} = \frac{q}{p {\cal S} (\rho)}$$ where ${\cal S} (\rho)$ is the occupation probability and $\rho=M/V$, the distribution function can be written, upto a 
normalization factor, as \be P(M) \propto e^{\sum_M [\ln (q/p {\cal 
S})]} \simeq e^{- V \int_0^{\rho} d\rho [\mu(\rho)-\mu_0]} 
\label{Pm_noncon} \ee where $\mu_0=\ln(q/p)$ an effective chemical 
potential and $f(\rho) = \int d\rho \mu(\rho)=\int d\rho \ln {\cal 
S} (\rho)$ an effective free energy (canonical) density function. 
The steady state mass density as a function of adsorption to 
desorption ratio $r = q/p$ can be obtained by minimizing the grand 
potential or the large deviation function for the density 
fluctuation $h(\rho) = f(\rho) - \mu_0 \rho$, leading to the 
relation ${\cal S} (\rho) = r$ (for details, see Appendix C).

Till now, the analysis is exact. However, it may not always be  
possible to exactly calculate the occupation probability ${\cal S} 
(\rho)$. For the purpose of demonstration, let us proceed by 
considering a model with diffusion and fragmentation rate as in 
Case I. We obtain an approximate expression, obtained within 
mean-field theory, of the occupation probability  (see Appendix C)
$${\cal S} (\rho) = \frac{\rho(1-\rho)}{(1+\rho)}.$$ Then, 
Eq. \ref{Pm_noncon} implies the subsystem 
mass distribution having a form $P_v(m) \propto w_v(m) \exp(\mu m)$ 
and consequently a FR relation as in Eq. \ref{FR1} follows. Then, for 
$\rho<\rho_c$  or equivalently for $r < r_c$, one can immediately 
calculate the scaled variance as 
\be \sigma^2 (\rho) = \left( \frac{d\mu_0}{d\rho} \right)^{-1} 
= \frac{\rho(1-\rho) (1+\rho)}{(1-2\rho-\rho^2)}, 
\ee 
where critical density $\rho_c=\sqrt{2}-1$.  
The variance in nonconserved case is 
different from that in the conserved-mass case, implying that the 
canonical and grand canonical ensembles are not equivalent
\cite{Mukamel_PRL2012, PRE2015}. However, the nature of 
singularity in the variance remains the same near criticality where 
$\sigma^2(\rho) \sim (\rho_c-\rho)^{-n}$ with $n=1$. Therefore the 
additivity property leads to the same power law scaling in the 
single-site mass distribution $P_1(m) \sim m^{-\tau} \exp(\tilde 
\mu m)$, for large $m$, where $\tau=5/2$ and $\tilde \mu = \mu_0 - 
\ln {\cal S} (\rho_c) = \ln (r/r_c)$ with $r_c = {\cal S} 
(\rho_c)$.

The above results are consistent with what was found, on the 
mean-field level, for general $p$ and $q$ in {\it `in-out'} 
model \cite{Barma_PRE2000} - a special case of the above
nonconserved model with $w=0$. One can interpret the 
results in the light of equilibrium BEC: The critical density 
signifies that, for $r > r_c = {\cal S}(\rho_c)= 3-2\sqrt{2}$, 
there is a condensate as in the BEC. In the grand-canonical 
setting (i.e., with no mass conservation), that would imply 
a phase with a diverging mass density, similar to the `Takayasu 
phase' where mass density actually diverges. For $p$ and $q$ 
finite, form of the subsystem mass distribution as written in Eq. 
\ref{Pm_noncon} remains the same, but only that the expression of 
${\cal S} (\rho)$, due to the presence of spatial correlations, is 
different. However, the similarity with the BEC still persists.

\section{Summary and Concluding Perspective}

In this paper, we argue that an additivity property can possibly explain why simple power-law scaling appears generically in 
nonequilibrium steady states with short-ranged correlations. We demonstrate that the existence of a fluctuation-response relation, a direct consequence of additivity, with a singular response function leads to power-law distributions with nontrivial exponents. The simplest form of the singularity, a simple pole, gives rise 
to the exponent $5/2$, which was often observed in the past in 
apparently unrelated systems. We substantiate the claims by 
analytically calculating the response function, which diverges as 
critical point is approached, in paradigmatic nonequilibrium mass 
aggregation models and the corresponding single-site as well as subsystem mass distributions. 
Most remarkably, the analysis, being independent of dynamical rules 
in a particular system, equally extends to critical properties in 
equilibrium and nonequilibrium.

Thermodynamic characterization of phase coexistence in driven 
systems is a fundamental problem in statistical physics and has been addressed in the past \cite{Bertin_PRL2006, Bertin_PRE2007, Majumdar_PRL2005, Majumdar_JPhysA2004, Majumdar_JSTAT2004, PRE2011, Dickman_PRE2015}, either numerically or analytically only for exactly known steady-states mostly having a product measure. 
From that perspective, it is quite encouraging that, even when steady-state weights are a priori {\it not known}, our analytical method not only gives insights into the steady-state structure but can also be applied to identify a chemical potential, which equalizes in the coexisting phases and vanishing of which at the criticality gives rise to pure power laws.

Note that, in our formulation, the mass distribution functions, though approximate, have been calculated solely from the knowledge of the variance. This formulation is perhaps not surprising in equilibrium where free energy function (or entropy, for an isolated system) essentially determines fluctuation properties of a system. However, in nonequilibrium scenario, it is a-priori not clear that such equilibrium thermodynamic approach can indeed be applied in systems having a steady state with nontrivial spatial structure. Here, it is worth mentioning that one requires, in principle, all the moments to specify a probability distribution function.
However, additivity property, provided it holds, puts a strong constraint on the mass distribution function $P_v(m)$ through a 
fluctuation-response relation and thus helps to uniquely determine $P_v(m),$ only from the knowledge of the variance as a function of density.

We believe that our analysis, being based on a general thermodynamic principle, would be applicable in many other driven systems where phase coexistence is known to occur (e.g., in active matters \cite{active_matter, Subhadip}). As a concluding remark, we mention that additivity property is expected to be quite generic for systems having short-ranged correlations and, therefore, it would be indeed interesting to actually verify additivity, through the predictions concerning fluctuations, on a case-by-case basis. Also, it remains to be seen whether the principle of additivity can be extended to systems having long-ranged spatial correlations, at least in the cases where the strength of these correlations are weak.

\section{Acknowledgement}

We thank R. Rajesh for helpful discussions 
and critical comments on the manuscripts. S.C. acknowledges the 
financial support from the Council of Scientific and Industrial 
Research, India [09/575(0099)/2012-EMR-I]. P.P. and P.K.M. acknowledge the financial support from the Science and Engineering Research Board (Grant No. EMR/2014/000719).

\section*{APPENDIX}

In this Appendix, we provide the details of the 
calculations to obtain the mass distributions, using additivity 
property, in mass aggregation models (both conserved and 
nonconserved versions) which were studied over the last couple of 
decades. The generalized models introduced here cover some of those 
studied in the past and their variants  \cite{Vigil, Krapivsky, 
Barma_PRL1998, Barma_JSP2000, Barma_PRE2000, Rajesh_PRE2002, Puri}

\section*{APPENDIX A: CALCULATION OF VARIANCE IN CONSERVED MASS AGGREGATION MODELS (CMAM)}

We define here a class of conserved mass aggregation models (CMAM) 
on a one dimensional lattice with periodic boundary and calculate 
the variance of mass at a single site in the steady state, assuming 
that the additivity property (Eq. 1) holds.  
For, simplicity, we consider only the discrete-mass cases. 

The mass at each site undergoes either diffusion (where whole of 
the mass is transferred to either of neighbouring sites) or 
chipping, with certain transition rates; in the models considered 
below, there are two types of chipping process. The diffusing mass 
or the chipped-off mass are coalesced with the mass at either of 
the neighbouring sites with a pre-assigned rates. In this process, 
the total mass of the system is conserved. 

Provided a site $i$ is occupied, particles hop to either of the 
nearest neighbour sites according to the transition rates specified 
below.
\\
\\
\textit{A. Diffusion with rate 1:} All particles at a site $i$ hop 
with rate 1 to left or right, i.e., $m_i \rightarrow 0$ and 
$m_{i\pm 1} \rightarrow m_{i\pm 1}+m_i$.\\ \\
\textit{B. Chipping with rate $w_1$:} This chipping process 
involves a particle at site $i$ being chipped off and thrown to 
left or right neighbour, i.e., $m_i \rightarrow (m_i - 1)$ and 
$m_{i\pm 1} \rightarrow m_{i\pm 1}+1$. \\ \\
\textit{C. Chipping with rate $w_2$:} This chipping process 
involves $m_i-1$ particles going to either left or right neighbour 
and the rest of the particles remaining at site $i$, i.e., $m_i 
\rightarrow 1$ and $m_{i\pm 1} \rightarrow m_{i\pm 1}+m_i-1$.
\\
\\
Assuming transition rates are Poissonian, we have the following stochastic update rules where mass $m_i(t+dt)$ at time $t+dt$ takes a particular value, depending on mass $m_i(t)$ at time $t$, with certain probabilities as shown below.
\\
\\
\textbf{Loss terms at site $i$:}
\\
\bea
m_i(t+dt)=\left\{
\begin{array}{ll} 
 \mbox{\underline{value:}} & \mbox{\underline{probability:}} \cr
 0 & dt \cr
 m_i(t) -1 + \delta_{m_i(t),0} & w_1 dt  \cr 
 1-\delta_{m_i(t),0} &  w_2 dt.
\end{array}
\right. \nonumber
\eea
\textbf{Gain terms from ${(i-1)}^{th}$ site:}
\bea
m_i(t+dt)=\left\{
\begin{array}{ll} 
 \mbox{\underline{value:}} & \mbox{\underline{prob.:}} \cr
 m_i(t)+m_{i-1}(t) & \frac {dt}{2}, \cr
 m_i(t) +1 -\delta_{m_{i-1}(t),0} & w_1 \frac{dt}{2},  \cr 
m_i(t)+m_{i-1}(t)-1+\delta_{m_{i-1}(t),0} & w_2 \frac{dt}{2}.
\end{array}
\right. \nonumber
\eea
\textbf{Gain terms from ${(i+1)}^{th}$ site:}
\bea
m_i(t+dt)=\left\{
\begin{array}{ll} 
\mbox{\underline{value:}} & \mbox{\underline{prob.:}} \cr
m_i(t)+m_{i+1}(t)  & \frac {dt}{2}, \cr
m_i(t)+1 -\delta_{m_{i+1}(t),0} & w_1\frac {dt}{2} \cr 
 m_i(t)+m_{i+1}(t)-1+\delta_{m_{i+1}(t),0} &  w_2\frac {dt}{2}.
\end{array}
\right. \nonumber
\eea
\textbf{Mass remains unchanged at site $i$:}
\bea
m_i(t+dt) =\left\{
\begin{array}{cc} 
\mbox{\underline{value:}} & \mbox{\underline{prob.:}} \cr
 m_i(t) & ~~~~(1-2dt-2w_1dt-2w_2dt). 
\end{array}
\right.
\nonumber
\eea
Now we define the occupation probability $\langle (1- \delta_{m_j,0}) \rangle = {\cal S}(\rho)$, i.e., the probability that a site is occupied. We deal with steady-state averages
throughout. We assume  that the additivity property (as in Eq. 1) is valid at single site level and therefore 
$n$-point ($n \ge 2$) correlation factorizes. 
\\
\\
\textbf{\bf{$n$-th} moment equation:} The time evolution of $n$-th 
moment $\langle m_i^n \rangle$ can be written as
\begin{widetext}
\bea
\langle m_i^n(t+dt) \rangle   
=\langle m_i^n(t) \rangle  
= \langle [m_i(t) -1 + \delta_{m_i(t),0}]^n \rangle w_1 dt 
+ \langle [m_i(t)+m_{i-1}(t)]^n \rangle \frac{dt}{2} 
+ \langle [1-\delta_{m_i(t),0}]^n \rangle w_2 dt 
\nonumber \\
+  \langle [m_i(t) +1 -\delta_{m_{i-1}(t),0}]^n \rangle w_1  \frac{dt}{2}
+ \langle [m_i(t)+m_{i-1}(t)-1+\delta_{m_{i-1}(t),0}]^n \rangle w_2 \frac{dt}{2} 
+  \langle [m_i(t) + m_{i+1}(t)]^n \rangle \frac{dt}{2} 
\nonumber \\
+ \langle [m_i(t)+1 -\delta_{m_{i+1}(t),0}]^n \rangle w_1 \frac{dt}{2} 
+ \langle [m_i(t)+m_{i+1}(t)-1+\delta_{m_{i+1}(t),0}]^n \rangle   w_2 \frac{dt}{2} 
\nonumber \\
+ \langle m_i^n(t) \rangle(1-2 dt-2 w_1 dt-2 w_2 dt),
\eea 
\end{widetext}
which, in the steady state where $\langle m_i^n(t+dt) \rangle = 
\langle m_i^n(t) \rangle$, gives a BBGKY hierarchy where $n$-point correlations are coupled to $(n+1)$-point correlations. 
To get a closed
set of equations for the moments, we use the factorization property 
of $n$-point correlations. As mentioned in the paper, the mass 
distributions are {\it solely} obtained from the response function 
(or the variance of the mass distribution) and therefore we are 
interested in only calculating the variance, or equivalently the 
second moment, which can be done as follows.
\\
\\
\textbf{$\bf{2^{nd}}$ moment equation:} If we put $n=2$ in the 
above equation, the second moment $\langle m_i^2 \rangle$ however 
cancels out from the above equation. Using factorization of 
two-point correlation, i.e., $\langle m_i m_j \rangle \approx 
\rho^2$ for $i \ne j$, we get an expression for the occupation 
probability $\cal S(\rho)$ as a function of mass density $\rho$,
\be
\rho^2 (1 + w_2) = w_+(\rho - {\cal S}) - w_- \rho 
{\cal S}, 
\ee
where   $w_{\pm} = w_1  \pm  w_2.$  This  gives
\be
{\cal S}(\rho) = \frac{w_+\rho - (1+w_2)\rho^2}{w_+ +w_-\rho}. \label{s_general}
\ee
\\
\\
\textbf{$\bf{3^{rd}}$ moment equation:} Similarly, for $n=3$, we 
obtain an equation where the third moment $\langle m_i^3 \rangle$ 
cancels out and we actually get, using factorization of both 
two-point and three-point correlation, a relation for the second 
moment
\be
\langle m^2 \rangle = \rho \frac{w_+(1+{\cal S})-2w_2\rho}{w_+-2(1+w_2)\rho - w_- {\cal S}}\label{av_msqr}
\ee
Using the expression of occupation probability in Eq. \ref{s_general}, we obtain
\be
\langle m^2 \rangle = \rho \frac{w_+^2+2w_+w_- \rho -(w_++3w_1w_2-w_2^2)\rho^2}{w_+^2-2w_+(1+w_2)\rho - w_-(1+w_2)\rho^2}
\ee
which leads to the desired expression of the variance as a function of density,
\bea
\sigma^2(\rho) &=&  \frac{w_+^2 \rho +w_+(w_1-3w_2)\rho^2}{w_+^2-2w_+(1+w_2)\rho - w_-(1+w_2)\rho^2} \cr
&+& \frac{(w_+-w_1w_2+3w_2^2)\rho^3+w_-(1+w_2)\rho^4}{w_+^2-2w_+(1+w_2)\rho - w_-(1+w_2)\rho^2}.
\label{sigma_GCMAM}
\eea
The variance $\sigma^2(\rho)$ has a {\it singularity} at $\rho = 
\rho_c$, i.e., it diverges at a critical density $\rho=\rho_c$,
which can be obtained by putting the denominator of Eq. \ref{sigma_GCMAM} zero and solving
\be
w_+^2-2w_+(1+w_2)\rho_c - w_-(1+w_2)\rho_c^2 = 0.
\ee
This gives a {\it simple pole} at the critical density
\be
\rho_c = {\frac{w_+}{w_-}} \left( \sqrt{1+\frac{w_-}{1+w_2}}-1 \right) .
\ee
Nonequilibrium free energy function can be calculated by 
integrating nonequilibrium chemical potential w.r.t. density 
$\rho$,
\be 
\mu(\rho)=\frac{d f}{d\rho}  \Rightarrow f(\rho) = \int \mu(\rho) d \rho \label{mu1}.
\ee
The function $\lambda_v(s) = - \ln \tilde{w}(s)$, which is the 
Legendre transform of the free energy density $f(\rho)$, can be 
obtained as given below, \be \lambda_v(s)  = v [f(\rho^*)+s
\rho^*], \ee
where $\rho^*$ is the solution of  
\be
s = -\mu(\rho^*). \label{s1_A} 
\ee

\section*{APPENDIX B: CALCULATION OF MASS DISTRIBUTION IN THE CONSERVED MASS AGGREGATION MODEL}

Here we provide the essential steps of the calculations to obtain 
single-site (i.e., $v=1$) mass distribution function $P_1(m) 
\propto w_1(m) \exp[\mu(\rho) m]$ where $w_1(m)$ is the single-site 
weight factor and $\mu(\rho)$ is a chemical potential. We first 
analyse the behaviour of $\lambda_1(s)$ near the singularity at
$s=s_c$ by expanding $\mu(\rho)$ and $f(\rho)$ near critical
density in the power series of $\rho-\rho_c$ where $\rho-\rho_c<0$
is small, 
\bea \mu(\rho)&=& \mu(\rho_c) + \frac{\mu''(\rho_c)}{2}
(\rho-\rho_c)^2 + \dots \label{mu2} \\
f(\rho) &=& f(\rho_c) + \mu(\rho_c) (\rho-\rho_c) +
\frac{f'''(\rho_c)}{3!} (\rho-\rho_c)^3 + \dots  \nonumber \eea
where we have
used Eq. \ref{mu1} and $\mu'(\rho_c)=f''(\rho_c)=0$ (see Fig.
\ref{mu-f-l}). Using Eq. \ref{s1_A} in Eq. \ref{mu2} and substituting $s +
\mu(\rho_c) \simeq - \mu''(\rho_c) (\rho^*-\rho_c)^2/2$, we get
\be (\rho^* - \rho_c) = - \sqrt{\frac{2}{|\mu''(\rho_c)|}}
(s-s_c)^{1/2} \ee where $s_c = - \mu(\rho_c)$ and
$\mu''(\rho_c)<0$. Therefore $\lambda_1(s)=f(\rho^*)+s\rho^*$ near
$s = s_c$, in the leading order of $(s - s_c)$, can be
approximated as \bea \lambda_1(s) &\simeq & \left[f(\rho_c) - 
s_c(\rho^*-\rho_c) + \frac{f'''(\rho_c)}{3!} (\rho^*-\rho_c)^3 
\right] + s\rho^* \nonumber \\ & = & \lambda_1(s_c) + \rho^*(s-s_c) 
+ \frac{f'''(\rho_c)}{3!} (\rho^* - \rho_c)^3 \nonumber \\
& = & \left[ a_0 + a_1 (s-s_c) + a_2 (s-s_c)^{3/2} \right] \eea where
$a_0=\lambda_1(s_c)=f(\rho_c)+s_c \rho_c$, $a_1=\rho_c$ and $a_2=-
(2/3)\sqrt{2/|\mu''(\rho_c)|}$. The inverse Laplace transform of 
the weight factor $w_1(m)$ can be written as
\be
\tilde w_1(s) = e^{- \lambda_1(s)} \simeq e^{-a_0} [1 - a_1 (s-s_c) - a_2 (s-s_c)^{3/2}] 
\ee
which, for $m \gg 1$, translates into 
\be
w_1(m) \sim \frac{e^{s_c m}}{m^{5/2}}.
\ee
Consequently the mass distribution can be written as
\bea
P_1(m) \sim \frac{e^{s_c m}}{m^{5/2}} e^{\mu(\rho) m} = \frac{e^{-(\alpha+\mu_0(\rho_c)) m}}{m^{5/2}} e^{(\mu_0(\rho) +\alpha) m} 
\\
P_1(m) \sim \frac{1}{m^{5/2}} e^{[\mu_0(\rho)-\mu_0(\rho_c)] m}.
\label{Pm} 
\eea
Note that effective chemical potential $\tilde \mu (\rho) = 
\mu_0(\rho)-\mu_0(\rho_c)$ is zero at the critical density 
$\rho_c = (\sqrt{2} - 1)$. The mass distribution in Eq. \ref{Pm} is 
precisely what was found in \cite{Barma_PRL1998} at $\rho = \rho_c 
$ and describes the simulation data remarkably well (see Fig. 1).

\section*{APPENDIX C: CALCULATION OF MASS DISTRIBUTION IN THE ABSENCE OF MASS CONSERVATION}

As shown in the paper, the probability distribution function $P(M)$ of total mass $M$ can be written, up to a normalization factor, as
\be P(M) = {\rm const.} \times e^{- V \int_0^{\rho} d\rho 
[\mu(\rho)-\mu_0]} \label{Pm_noncon_AP} \ee
Now, if we assume that the joint mass distribution ${\cal P} [\{m_i\}]$ has a product form on single-site level ($v=1$), i.e., product of single-site mass distribution function $p(m_i)$,
\be 
{\cal P} [\{m_i\}] = \prod_{i=1}^V p(m_i),
\ee
the probability distribution function $P(M)$ of mass $M$ in the system can be written as
\be
P(M) = \prod_{i=1}^V \left[ \int dm_i p(m_i) \right] \delta \left( M - \sum_i m_i \right).
\ee
From the Laplace transform $\tilde P (s) = \int dM P(M) \exp(-s M) = [\tilde p(s)]^V$ of the mass distribution $P(M)$, the Laplace transform $\tilde p(s) = \int dm_i p(m_i) \exp(-s m_i)$ of 
single-site mass distribution $p(m)$ can be written as 
\be
\tilde p (s) = {\rm const.} \times e^{-\lambda_1(s)},
\ee
where 
\be
\lambda_1(s) = {\rm \bf inf}_\rho [h(\rho) + s\rho].
\ee
Here we have used inverse transform
\be
\tilde P (s) = {\rm const.} \times \int d\rho e^{-V[h(\rho) + s\rho]},
\ee
which has been obtained from Eq. \ref{Pm_noncon_AP} and where grand 
potential or the large deviation function for density fluctuation 
$h(\rho) = f(\rho) - \mu_0 \rho = \int_0^{\rho} [\mu(\rho) - \mu_0 ] d\rho$ and chemical potential $\mu(\rho) = \ln {\cal S} (\rho) = 
\ln [\rho(1-\rho)/(1+\rho)]$, as given in the paper. Note that 
the function ${\cal S} (\rho)$ is the occupation probability in the 
conserved mass aggregation model and has been obtained by putting 
$w_1=1$ and $w_2=0$ in Eq. \ref{s_general}.

Now the function $\lambda_1(s)$, Legendre transform of grand potential $h(\rho)$, can be written as
\be
\lambda_1(s) = h(\rho^*) + s \rho^*,
\ee
where $\rho^*$ is the root of the equation $d[h(\rho)+s \rho]/d\rho = 0$ or $\mu(\rho^*)-\mu_0 + s = 0$, i.e., $\rho^*$ is the root of 
\be 
\ln \left[ \frac{\rho^*(1-\rho^*)}{1+\rho^*} \right] = \mu_0 - s.
\ee 
The critical density is obtained by putting scaled variance as $\sigma^2(\rho) = (d\mu_0/d\rho)^{-1} = \infty$ or $1/\sigma^2(\rho) = 0$,
\be 
\frac{(1-2\rho_c - \rho_c^2)}{\rho_c(1 - \rho_c)(1 + \rho_c)} = 0,
\ee
and thus $\rho_c = \sqrt{2} - 1$. In the macrostate (most probable 
state), we have ${\cal S}(\rho)=r$, implying that the critical 
density is related to the ratio $r = q/p$ through ${\cal S} 
(\rho_c) = r_c$. To obtain the large-mass behaviour, we 
expand $\mu(\rho)$ around $\rho=\rho_c$,
\be 
\mu(\rho) = \mu(\rho_c) + \frac{\mu''(\rho_c)}{2} (\rho - \rho_c)^2,
\ee
to obtain 
\bea
(s-s_c) \simeq \frac{|\mu''(\rho_c)|}{2} (\rho^* - \rho_c)^2, \\
\lambda_1(s) \simeq a_0 + a_1 (s-s_c) + a_2 (s-s_c)^{3/2},
\eea
in leading order in $(\rho^* - \rho_c)$ where $s_c = \mu_0 - \mu(\rho_c)$, leading to the desired result in the paper, 
\be 
p(m) \sim \frac{1}{m^{5/2}} e^{s_c m} = \frac{1}{m^{5/2}} e^{[\mu_0 - \mu(\rho_c)] m}.
\ee


\begin{thebibliography}{99}
\bibitem{note} For example, a normalizable distribution $P_v(m) \sim \frac{1}{m_1^*} \Phi_1 ( \frac{m}{m_1^*}) + \frac{1}{(m_2^*)^2} \Phi_2( \frac{m}{m_2^*} )$, near criticality, can have a finite mean $\langle m \rangle =\rho$ but diverging second moment $\langle m^2 \rangle$ for a suitable choice of a finite $m_1^*(\rho)$ and a diverging $m_2^*(\rho) \sim (\rho - \rho_c)^{-n}$; a simple choice could be $\Phi_{1,2}(x) = \exp( - \kappa_{1,2} x^{\delta_{1,2}})$ with $\kappa_{1,2} , \delta_{1,2}>0.$
\end{thebibliography}
\end{document}